\documentstyle[prd,aps,floats]{revtex}

\begin{document}
\preprint{SUSSEX-AST 97/4-2, astro-ph/9704251}
\draft

%
%
\input epsf
\renewcommand{\topfraction}{0.8}
\twocolumn[\hsize\textwidth\columnwidth\hsize\csname 
@twocolumnfalse\endcsname

\title{Constraints on the density perturbation spectrum from primordial
 black holes} 
\author{Anne M.~Green and Andrew R.~Liddle}
\address{Astronomy Centre, University of Sussex, Falmer, Brighton BN1
9QH,~~~U.~K.}  
\date{\today} 
\maketitle
\begin{abstract}
We re-examine the constraints on the density perturbation spectrum, 
including its spectral index $n$, from the production of primordial black 
holes. The standard cosmology, where the Universe is radiation dominated 
from the end of inflation up until the recent past, was studied by Carr, 
Gilbert and Lidsey; we correct two errors in their derivation and find a 
significantly stronger constraint than they did, $n \lesssim 1.25$ rather 
than their 1.5. We then consider an alternative cosmology in which a 
second 
period of inflation, known as thermal inflation and designed to solve 
additional relic over-density problems, occurs at a lower energy scale 
than 
the main inflationary period. In that case, the constraint weakens to $n 
\lesssim 1.3$, and thermal inflation also leads to a `missing mass' range, 
$10^{18} \, {\rm g} \lesssim M \lesssim 10^{26} \, {\rm g}$, in which 
primordial black holes cannot form. Finally, we discuss the effect of 
allowing for the expected non-gaussianity in the density perturbations 
predicted by Bullock and Primack, which can weaken the constraints further 
by up to 0.05. 
\end{abstract}

\pacs{PACS numbers: 98.80.Cq \hspace*{1.9cm} Sussex preprint SUSSEX-AST 
97/4-2, astro-ph/9704251}

\vskip2pc]

\section{Introduction}

Primordial black holes (PBHs) are formed in the early universe if
density perturbations are sufficiently large, and provide a useful
probe of the primordial power spectrum over a wide range of
scales. The data from COBE and from large-scale structure observations
constrain the power spectrum on large scales (from about one megaparsec up 
to thousands of megaparsecs), whereas PBHs may form over a wide range of 
smaller scales (1~Mpc to $10^{-16}$~Mpc). Limits on their production can 
be 
used to constrain inflation models in which the perturbations grow as one 
moves to shorter scales, the so-called blue spectra.
 
There are a number of well-known limits, covering various mass ranges,
on the maximum allowed mass fraction of PBHs
\cite{var:lim,BCL:PBH,CGL:PBH} . Some
are imposed at the present epoch and some at earlier stages such as
nucleosynthesis. These constraints fall into two categories, those
from the effects of Hawking radiation and those from the gravitational
effects alone. The evaporation of PBHs via thermal emission has
potentially observable astrophysical consequences, and while no
unambiguous detection has been made, observations have placed limits
on the maximum mass fraction of PBHs allowed at evaporation. PBHs with
mass $M \lesssim 5 \times 10^{14}$~g will have evaporated before the
present epoch.  PBHs more massive than this will not have experienced
significant evaporation, and their present density must not overclose
the universe: $\Omega_{\rm{pbh},0} < 1$. An additional, less secure,
constraint on light PBHs can be obtained if one supposes that black
hole evaporation leaves a stable relic, normally assumed to have a
mass of order the Planck mass, rather than evaporating to nothing.

It is thought that PBHs can only form in the early universe, and that
they do so at a time when the horizon mass equals the black hole
mass. A number of formation mechanisms are possible 
\cite{var:form,Carr:rev},
the simplest being formation from large density perturbations. The
horizon mass in a radiation-dominated universe with temperature $T$ is
given by
\begin{equation}
\label{mhor}
M_{{\rm H}} \simeq 10^{18} \, {\rm g} \, \left( \frac{10^7 \, {\rm
	GeV}}{T} \right)^2 \,,
\end{equation}
from which we see that evaporating PBHs must form very early indeed in
the history of the universe, in particular long before the epoch of
nucleosynthesis ($T \sim 1 \, {\rm MeV}$) at which the standard big
bang evolution is well validated.

In the `standard' cosmology, the universe has been radiation dominated
ever since the end of the reheating period after a phase of inflation
at extremely high energies, which was responsible for the generation
of density perturbations. Under this assumption, the limits on the PBH
density can be extrapolated backwards to the time of formation to give
limits on the initial mass fraction of PBHs, $\beta_{\rm{i}}
=\rho_{\rm{pbh,i}}/\rho_{\rm{tot,i}}$ where $\rho_{\rm{pbh,i}}$ and
$\rho_{\rm{tot,i}}$ are the PBH and total energy densities
respectively, at the time at which the PBHs are formed. Substantial
work has been done under this assumption
\cite{BCL:PBH,CGL:PBH,CL:chaotic}.  However, there is no direct
evidence requiring that the universe be radiation dominated at the
high temperatures under consideration, and these limits can be
greatly altered if the evolution of the Universe is more
complex. In this paper we consider one of the most dramatic possible
changes to the early evolution --- the effect of a second period of
inflation.
  
A recent extension to the standard cosmology is {\em thermal
inflation} \cite{LS:TI}, a second period of inflation due to a scalar
field known as a flaton. This has nothing to do with the normal period
of inflation, which is still assumed to occur at some higher energy
scale to solve the flatness and horizon problems and to generate
density perturbations. Flaton fields, which are a consequence of
supersymmetric theories, have a vacuum expectation value (vev) $M \gg
10^3$~GeV, even though their mass $m$ is only of order the
supersymmetry scale, $\sim 10^2$ to $10^3$~GeV. Their potential is
therefore almost flat. In the early universe these fields are held at
zero by finite temperature effects, with false vacuum energy density
$V_{0} \sim m^2 M^2$. Once the temperature falls below $V_{0}^{1/4}$,
the false vacuum energy density dominates the thermal energy
density. This false vacuum energy begins to drive a new period of
inflation. This inflation continues until the temperature drops to $T
\sim m$, at which point thermal effects are no longer strong enough to
anchor the flaton in the false vacuum. The most popular possible
identity for flaton fields are the moduli fields in superstring
theory.

A modulus field with a vev of order the Planck mass $m_{{\rm Pl}}$ (as is 
expected if
the vev is non-zero) would produce many particles which would not
decay before nucleosynthesis and hence would destroy the standard
model of cosmology; this is the `moduli problem' \cite{var:mod}. It
cannot be solved in the same way as the monopole problem by invoking
an early epoch of inflation lasting upwards of $60$ $e$-foldings,
since the energy scale at the end of inflation is normally greater
than $10^{12}$~GeV and the moduli would be regenerated after inflation.
To avoid too many moduli being regenerated requires~\cite{moreTI}
\begin{equation}
V_{{\rm inf}}^{1/4} \leq (10^7 \; \mbox{to} \; 10^8) \, {\rm GeV}
	\, \left( \frac{1 \, {\rm GeV}}{T_{{\rm R}}} \right)^{1/4} \,,
\end{equation}
where $T_{{\rm R}}$ is the reheat temperature. A single period of
inflation at such a low energy scale would not be capable of producing
the observed density perturbations, but a period of thermal inflation
can solve this problem provided $M$ is within a range of several
orders of magnitude around $10^{12}$~GeV. Taking $M\sim 10^{12}$~GeV
gives $V_{0}^{1/4} \sim 10^{7}$~GeV so that around $\ln(10^7/10^3)
\sim 10$ $e$-foldings of thermal inflation occur, sufficient to dilute
the moduli existing before thermal inflation but small enough to not
affect the density perturbations generated during the first period of
inflation.\footnote{Originally the flaton was taken to be the GUT
Higgs \cite{LS:TI} so that $M \sim M_{\rm{GUT}} \sim 10^{16}$~GeV and
$V_{0}^{1/4} \sim 10^{10}$~GeV, leading to $15$ $e$-foldings of
inflation. However, successful nucleosynthesis requires $M \lesssim
10^{12}$~GeV and thermalization of a stable lightest supersymmetric
particle, if one exists, requires $M \lesssim 10^{10}$~GeV
\cite{moreTI,BCLP}.}

The main effect of the period of thermal inflation is to dilute the
density of PBHs, relative to radiation, by a factor of
$(a_{\rm{f}}/a_{\rm{i}})^{3} \sim (10^{4})^3$, where $a_{\rm{i}}$ and
$a_{\rm{f}}$ are the scale factors immediately before and after
thermal inflation. In addition, there are two more subtle changes from
the standard scenario. Firstly, for PBHs which form in the time
between the two periods of inflation, the comoving scale to which a
given mass of PBH corresponds is changed. During thermal inflation the
horizon mass remains constant whilst the comoving Hubble radius grows,
so a given PBH mass corresponds to a shorter comoving length scale and
hence to a later stage in the original, density perturbation
generating, epoch of inflation. Secondly, thermal inflation introduces
a missing mass range of PBHs. This corresponds to those comoving
scales which enter the horizon before thermal inflation (possibly
forming black holes as they do so), and are then pulled back outside
again during thermal inflation. Any new density perturbations might be
expected to be small, since the energy scale of thermal inflation is
much lower than the original inflationary period, and hence unable to
form black holes when they re-enter the horizon again after thermal
inflation.\footnote{Note however the standard perturbation calculation
breaks down as the hypothesis that the initial state is the vacuum
probably cannot be justified. It is not clear how to make the
necessary generalization.} {}From Eq.~(\ref{mhor}), this corresponds to
masses in the range $10^{18} \, {\rm g} \lesssim M \lesssim 10^{26} \,
{\rm g}$.

Finally, PBH formation requires $\delta \rho/\rho$ on the
relevant scales to be two or three orders of magnitude larger than the 
value
required by COBE, which applies at very large scales.  PBHs can
therefore only be formed in significant numbers if the spectral index
$n$ of the density perturbations is significantly above unity,
normally referred to as a blue spectrum. Limits on PBH formation allow
upper bounds to be placed on $n$; for the standard cosmology this was
done by Carr et al.~\cite{CGL:PBH}. We shall first re-examine 
their calculation, and then generalize it to include the possibility of 
thermal inflation.

\section{Primordial Black Hole Formation}

In order for a PBH be formed, a collapsing overdense region must be
large enough to overcome the pressure force resisting its collapse as
it falls within its Schwarzschild radius. Consider a spherically
symmetric region with density $\tilde{\rho}$ greater than that of the
background, whose evolution will be governed by the positive curvature
Friedmann equation \cite{BP:gauss}. The perturbed region stops
expanding when $\tilde{H}=0$, at which time the region has size
$R_{\rm{c}} \approx \delta_{\rm{i}}^{-1/2} R_{\rm{i}}$ where
$R_{\rm{i}}$ is its size at some arbitrary initial time and
$\delta_{\rm{i}} = (\tilde{\rho_{\rm{i}}}
-\rho_{\rm{i}})/\rho_{\rm{i}}$ is the initial density perturbation
\cite{Har:form}. If the perturbed region contains enough matter to
overcome any pressure forces, it will continue to contract. This
requires that its radius exceeds the Jeans length, $R_{\rm{c}} \geq
R_{\rm{Jeans}}= c_{\rm{s}} t_{\rm{c}}$, where $c_{{\rm s}}$ is the
sound speed. In a radiation-dominated universe, $R_{{\rm Jeans}} \sim
t_{\rm{c}}/\sqrt{3}$. Now $R_{\rm{c}}/t_{\rm{c}} \approx
R_{\rm{i}}/t_{\rm{i}} \delta_{\rm{i}}^{-1/2} $, and since this
expression is constant with time we can evaluate it at horizon
crossing ($R=t$) leading to a constraint on the perturbations at
horizon crossing $\delta \geq 1/3$. There is also an upper limit of
$\delta \leq 1$; a perturbation which exceeded this value would
correspond initially to a separate closed universe 
\cite{Carr:delta,Carr:rev}, which is an inconsistent initial
condition for our purposes. So for PBH formation we require the
initial fluctuations to satisfy
\begin{equation}
\label{delrange}
1/3 \leq \delta \leq 1 \,.
\end{equation}

When a perturbation satisfying the above condition crosses the
horizon, a PBH will be formed with mass \cite{Carr:rev}
\begin{equation}
\label{M}
M = \gamma^{3/2} M_{{\rm H}} = \frac{\gamma^{3/2}}{g_{\star
	\rm{form}}^{1/2}} \left( \frac{t}{t_{{\rm Pl}}} \right) 
	m_{{\rm Pl}} \,,
\end{equation}
where the background equation of state is $p = \gamma \rho$,
$\gamma$ being $1/3$ in a radiation-dominated universe. Here 
\begin{equation}
\label{mhordef}
M_{\rm{H}}= \frac{4 \pi}{3} \rho (H^{-1})^{3} \,,
\end{equation}
and $g_{\star}$ is the effective number of massless degrees of freedom
at this time. We define $\beta_{\rm{i}}$ to be the initial mass
fraction of PBHs, which is given by the fraction of the Universe
satisfying Eq.~(\ref{delrange}):
\begin{equation}
\beta_{i} \equiv \frac{\rho_{\rm{pbh,i}}}{\rho_{\rm{tot,i}}} = 
	\int_{1/3}^{1} p(\delta) \, {\rm d} \delta \,,
\end{equation}
where $p(\delta)$ is the probability distribution for $\delta$.

Normally (for instance in large-scale structure studies) when one 
considers 
perturbations the probability distribution is assumed to be gaussian. This 
is well justified when the perturbations are small. However, Bullock and 
Primack \cite{BP:gauss} have recently challenged this assumption for PBH 
formation, since the perturbations cannot be very small if a significant 
formation rate is to be obtained. They typically find a suppression of 
large 
perturbations relative to the gaussian hypothesis. While this suppression 
can be very dramatic when expressed in terms of the number of black holes 
formed, it actually does not lead to a large change in constraints on the 
perturbation spectrum. We shall therefore maintain the gaussian assumption 
for our derivations, and in Section~\ref{s:index} we shall assess the 
changes non-gaussianity introduces.

In order to examine specific mass ranges, we have to smooth
the density distribution, which is done in the normal way using a
window function $W(kR)$, which we take to be a top-hat. For gaussian
distributed fluctuations, the probability distribution of the smoothed
density field, $p(\delta(M))$, is given by
\begin{equation}
\label{gauss}
p(\delta(M)) \, {\rm d} \delta(M) = \frac{1}{ \sqrt{2 \pi} \sigma(M)} 
	\exp{\left( - \frac{\delta^2(M)}{2 \sigma^2(M)}\right)} \, 
	{\rm d} \delta(M) \,.
\end{equation}
Here $\sigma(M)$ is the mass variance evaluated at horizon crossing, 
defined for example in Ref.~\cite{LLrep} 
\begin{equation}
\label{sigmam}
\sigma^2(M) = \frac{1}{2 \pi^2} \int_{0}^{\infty} P(k) W^2(kR) k^2 
	\,{\rm d}k \,.
\end{equation}
where $P(k)=\langle|\delta_{k}|^{2}\rangle$ is the power spectrum,
the $\delta_{k}$ being the coefficients when $\delta(x)$ is Fourier
expanded. The power spectrum is usually taken to have primordial form
$P(k) \propto k^n$ for simplicity; in general there is no reason to
expect inflation models to produce power-law spectra over as wide as
range of scales as we will need to consider, though there are models
which do.

The mass fraction of black holes is given from the above by 
\begin{equation}
\beta(M) = \int_{1/3}^{1} \frac{1}{\sqrt{2 \pi} \, \sigma(M)} \exp
	{\left( - \frac{ \delta^2}{2 \sigma^2(M)} \right) } \, {\rm d}
	\delta \,.
\end{equation}
Since the integrand is a rapidly falling function in the regime of
interest, dropping by a factor $\exp(-0.5)$ every time $\delta$ is
increased by $\sigma$, this integral can be approximated by evaluating
the integrand at $\delta= 1/3$ and multiplying by $\sigma(M)$, leading
to
\begin{equation}
\label{beta}
\beta(M) \approx \sigma(M) \exp{ \left( - \frac{1}{18 \, \sigma^2(M)}
	\right)} \,.
\end{equation}
Strictly speaking, this is the mass fraction in
black holes of mass {\em greater than} $M$, but in practice $\beta(M)$
is such a rapidly falling function that these can be taken to all have
the same mass $M$.

\typeout{IF YOU GET A LaTeX ERROR HERE, READ THE COMMENT AT THE BOTTOM
OF THE MAIN LaTeX FILE. OTHERWISE IGNORE THIS! ANDREW}

\begin{table*}
\begin{center}
\begin{tabular}{|c|c|c|}
Constraint & Range & Reason \\ 
\hline 
\hline 
$\alpha_{\rm{evap}} < 0.04$ & $10^{9}$ g $< M < 10^{13}$ g & 
Entropy per baryon at nucleosynthesis \\ 
\hline 
$\alpha_{\rm{evap}} < 10^{-26} (M/m_{{\rm Pl}})$ & 
$M = 5\times10^{14}$~g & $\gamma$ rays
from current explosions \\ 
\hline $\alpha_{\rm{evap}} < 6\times10^{-10} (M/m_{{\rm Pl}})^{1/2}$
 & $10^{9}$~g $ < M <10^{11}$~g & n$\bar{\rm{n}}$ production at 
nucleosynthesis \\ 
\hline
$\alpha_{\rm{evap}} < 5\times10^{-29} (M/m_{{\rm Pl}})^{3/2}$ &
$10^{10}$~g $< M < 10^{11}$~g & Deuterium destruction \\ 
\hline
$\alpha_{\rm{evap}} < 1\times10^{-59}(M/m_{{\rm Pl}})^{7/2}$ &
$10^{11}$~g $< M < 10^{13}$~g & Helium-4 spallation
\end{tabular}
\end{center}
\caption[massfrac]{\label{massfrac} Limits on the mass fraction of PBHs at
evaporation.}
\end{table*}

The final step is to relate the mass scales to comoving scales during
inflation. In the notation of Ref.~\cite{LLrep}, the initial spectrum
of perturbations is $\delta_{\rm{H}}^2 (k) \propto k^{n-1}$ where
\begin{equation}
\delta_{\rm{H}}^2 = \left( \frac{k^3}{2 \pi^2} \right) \,
	\left( \frac{a H}{k} \right)^{4} \, P(k) \,;
\end{equation}
The quantity $\delta_{{\rm H}}$ stays constant on scales above the
Hubble radius and is a good estimate of the rms density contrast at
horizon entry~\cite{LLrep}. During radiation domination the comoving
Hubble radius $H^{-1}/a$ is proportional to $T^{-1}$, so that a given
scale $k$ crosses within the Hubble radius when
\begin{equation}
\label{hub}
k^{-1} = \frac{H^{-1}}{a} = \left( \frac{H^{-1}}{a} \right)_{{\rm eq}}
	\left( \frac{T_{{\rm eq}}}{T} \right) \,,
\end{equation}
where subscript `eq' refers to quantities evaluated at
matter-radiation equality. Meanwhile the horizon mass $M_{\rm{H}}$,
Eq.~(\ref{mhordef}), varies as $T^{-2}$ so that, in the absence of
thermal inflation, we get
\begin{equation}
\label{k}
k^{-1} =  \left( \frac{H^{-1}}{a} \right)_{{\rm eq}} \left( 
	\frac{M_{\rm{H}}}{M_{\rm{H,{\rm eq}}}} \right)^{1/2} \,.
\end{equation}
Substituting this into the expression for $\delta_{\rm{H}}(k)$ we
obtain 
\begin{equation}
\label{sigma}
\sigma_{{\rm hor}}(M) = \sigma_{{\rm hor}}(M_{\rm eq}) \left( 
	\frac{M}{M_{{\rm eq}}}\right)^{(1-n)/4} \,,
\end{equation}
where $M_{{\rm eq}}$ is the horizon mass at matter--radiation equality. We 
stress that this
equation refers to the dispersion at horizon crossing, not at constant
time.\footnote{ This disagrees with
Ref.~\cite{CGL:PBH}, in which a different scaling $\sigma(M) \propto
M^{(1-n)/6}$ was used. This arises from assuming $M \propto k^{-3}$
with no time-dependence --- i.e.~that the
{\em comoving} mass density is conserved. This is true for matter
domination but not for radiation domination where the comoving mass
density decreases with time. Because our scaling is stronger with $n$,
our final constraints on $n$ are tighter than theirs.}

The lightest black holes to form are those which enter the horizon 
immediately after inflation. For simplicity we shall assume prompt 
reheating, and Eq.~(\ref{mhor}) then gives the minimum mass. Carr et 
al.~\cite{CGL:PBH} examine some consequences of delayed reheating.

\section{Limits on the PBH abundance}

\label{limits}

\subsection{At evaporation}

The observational constraints on the mass fraction of black holes at
evaporation, $\alpha_{\rm {evap}}(M) = \rho_{\rm{pbh}} /
\rho_{\rm{rad}}$ are well known \cite{var:lim,BCL:PBH,CGL:PBH} , and
are listed in Table \ref{massfrac}.

To interpret these we need to relate the black hole mass to their
lifetime.  Carr \cite{Carr:rev} parametrizes the results of Page
\cite{Page:tevap}, which were found numerically by considering the
number of species which a black hole of given mass can emit at a
significant rate, to give the following relation between PBH mass and
lifetime 
\begin{equation}
\tau_{\rm{evap}} = \frac{9 \times 10^{-27}}{f(M)} \left( 
	\frac{M}{1\,\rm{g}}\right)^3 {\rm sec} \,,
\end{equation}
where $f(M)$ depends on the number of particle species which can be
emitted and is normalized to 1 for holes which emit only massless
particles. Note that the bulk of the evaporation always takes place
near the initial temperature, so in this expression one only needs $f$
at the initial mass and not as a time-varying quantity.  Considering
the number of spin states available for a PBH to evaporate into at the
present day ($g_{\rm{eff,0}})$; there are two polarizations of photon
plus three neutrino species which each have two spin states and give a
contribution 7/8 times that of the photons, since they obey Fermi
rather than Bose statistics.  Therefore $g_{\rm{eff,0}}= 2 + (3 \times
2 \times 7/8) = 7.25$ so that $f(M) \equiv g_{\rm{eff,0}}/7.25$ and
\begin{equation}
\label{tevap}
\tau_{\rm{evap}} = \frac{1.2 \times 10^{4}}{g_{\rm{eff}}} \left( 
	\frac{M}{m_{{\rm Pl}}} \right)^{3} t_{{\rm Pl}} \,,
\end{equation}
with the value of $g_{\rm{eff}} $ at the time of evaporation being
taken. For $M> 4 \times 10^{9}$~g the temperature at evaporation is
sufficiently low ($< 10^{-4} $GeV) that $g_{\rm{eff}}$ has its present
day value of 7.25.  Eq.~(\ref{tevap}) is often quoted without the
factor of $1.2 \times 10^{4}$; this is certainly non-negligible when
calculating PBH lifetime although other approximations often made when
limiting the initial PBH abundance (e.g.~$g_{\rm{eff}} \sim 1$, $\gamma
\sim 1$ and $M \sim M_{\rm{H}}$ in Ref.~\cite{CGL:PBH}) appear
to largely cancel this factor.

\subsection{Present-day PBH density}

{}From Eq.~(\ref{tevap}), PBHs of mass $M> 5\times 10^{14}$~g will not
have evaporated by the present day, but their initial abundance can be
constrained from the fact that their present mass density must not
overclose the universe 
\begin{equation}
\Omega_{{\rm pbh,0}} = \Omega_{\rm{pbh,eq}} <1 \,.
\end{equation}

\subsection{Present-day relic density}

It has been argued \cite{var:rel} that PBHs may not evaporate
completely, as originally assumed, but instead leave a relic with mass
$M_{\rm{rel}} \sim m_{{\rm Pl}}$. If this is the case the present mass
density of relics, which will remain from all PBHs with initial mass
$M< 5 \times 10^{14}$~g, similarly must not overclose the universe,
leading to
\begin{eqnarray}
\label{rel}
\Omega_{\rm{rel,0}} & = & {\Omega}_{\rm{rel,eq}} =
	\left(\frac{m_{\rm{Pl}}}{M}\right)
	\frac{\rho_{\rm{pbh,i}}}{\rho_{\rm{tot,eq}}}
	\left(\frac{T_{\rm{0}}}{T_{\rm{form}}}\right)^3 \nonumber \\ &
	= & \left(\frac{m_{\rm{Pl}}}{M}\right)
	\frac{\beta_{\rm{i}}}{1-\beta_{\rm{i}}}
	\frac{\rho_{\rm{rad,i}}}{\rho_{\rm{tot,eq}}}
	\left(\frac{T_{\rm{0}}}{T_{\rm{form}}}\right)^3 <1 \,.
\end{eqnarray}

\section{Limits on initial mass fraction of PBHs}

\label{s:massfrac}

\subsection{Standard evolution of the universe}

 In constraining the initial mass function of the
black holes, one needs to assume an entire history for the universe
from the time of formation to the present. Traditionally, it has been
assumed that the universe was radiation dominated up until the recent
matter-dominated era, and then the limits of Section \ref{limits} can
easily be evolved backwards in time in order to constrain the initial
mass fraction of PBHs.  The energy density in radiation dilutes as
$\rho_{\rm{rad}} \propto a^{-4}$, whereas that in PBHs decreases more
slowly, $\rho_{\rm{pbh}} \propto a^{-3}$.  Therefore
\begin{equation}
\left( \frac{\rho_{\rm{pbh}}}{\rho_{\rm{rad}}} \right)_{\rm{evap}} =
	\alpha(M)_{\rm{evap}} = \frac{\beta_{\rm{i}}}{ 1-\beta_{\rm
	{i}}} \left( \frac{t_{\rm{evap}}}{t_{\rm{form}}}
	\right)^{1/2} \,.
\end{equation}
Using Eqs.~(\ref{M}) and (\ref{tevap}), and taking $g_{\star} \sim 100$
\begin{equation}
\alpha(M)_{\rm{evap}} = 3.2 \frac{\beta_{\rm{i}}}{ 1-\beta_{\rm{i}}}
	\left( \frac{M}{m_{{\rm Pl}}} \right) \,.
\end{equation}

\begin{table}
\begin{center}
\begin{tabular}{|c|c|}
Constraint & Range \\ \hline $ \alpha_{\rm{i}} < 3\times10^{-16} \,
(10^{9} \, {\rm g}/M)$ & $ 10^{9} \, {\rm g}< M < 10^{13} \, {\rm g}$ \\ 
\hline 
$ \alpha_{\rm{i}}
<3\times 10^{-27}$ & $M \simeq 5\times10^{14} \, {\rm g}$ \\ 
\hline 
$\alpha_{\rm{i}} <3\times 10^{-17} \,
(10^{9} \, {\rm g}/M)^{1/2} $ & $10^{9} \, {\rm g}< M < 10^{11} \, {\rm 
g}$ \\ 
\hline 
$\alpha_{\rm{i}} <3\times10^{-22} \,
(M/10^{10} \, {\rm g})^{1/2} $ & $10^{10} \, {\rm g}< M < 10^{11}\, {\rm 
g}$ \\ 
\hline 
$\alpha_{\rm{i}} <3\times 10^{-21} \,
(M/10^{9} \, {\rm g})^{5/2} $ & $10^{11} \, {\rm g}< M <10^{13}\, {\rm g}$ 
\\ 
\hline 
$ \alpha_{\rm{i}} < 1\times10^{-19} \,
(M/10^{15} \, {\rm g})^{1/2} $ & $M>10^{15} \, {\rm g}$ \\
\hline 
$\alpha_{\rm{i}} < 0.1 \, (M/10^{15} \, {\rm g})^{3/2} $ & $M<10^{15} \, 
{\rm g}$
\end{tabular}
\end{center}
\caption[initmass]{\label{initmass} Limits on the initial mass fraction of
PBHs, without thermal inflation. We define $\alpha_{\rm{i}} \equiv
\beta_{\rm{i}}/(1-\beta_{\rm{i}})$ for compactness.  The final
constraint is the relic constraint which only
applies if relics are assumed.}
\end{table}

The gravitational constraint can be evaluated simply:
\begin{eqnarray}
\rho_{\rm{PBH,eq}} &  = & \rho_{\rm{pbh,i}} \left( \frac{ 
	T_{\rm{eq}}}{T_{\rm{form}}} \right)^{3} \nonumber \\
	& = &\frac{\beta_{\rm{i}}}{1-\beta_{\rm{i}}} \frac{\pi^2}{30} 
	g_{\star}^{\rm{form}} T_{\rm{form}}^4 \left( 
	\frac{ T_{\rm{eq}}}{T_{\rm{form}}} \right)^{3} \,,
\end{eqnarray}
and
\begin{equation}
\rho_{\rm{tot,eq}} = 2 \frac{\pi^2}{30} g_{\star}^{\rm{eq}}
	T_{\rm{eq}}^4 \,,
\end{equation}
where $g_{\star}^{\rm{form}} \sim 100$ and $g_{\star}^{\rm{eq}} \sim
3$,\footnote{Note that $g_{\star}$ is the effective number of degrees
of freedom as far as cosmology is concerned, evaluated at the photon
temperature. It is lower than $g_{{\rm eff}}$ quoted earlier since the
cosmic neutrino background is at a lower temperature than the
microwave background.} so that
\begin{equation}
\Omega_{\rm{pbh,0}} = 17 \frac{\beta_{\rm{i}}}{1-\beta_{\rm{i}}} \left( 
	\frac{t_{\rm{form}}}{t_{\rm{eq}}} \right)^{1/2} \,.
\end{equation}
Taking $t_{{\rm eq}} = t_{0} \Omega_{\rm{rel}}^{3/2} = 6.5 h^{-1}
\rm{Gyr} \times (4 \times 10^{-5} h^{-2} )^{3/2}$ and using
Eq.~(\ref{M})
\begin{equation}
\Omega_{{\rm pbh,0}} = 6.7 \times 10^{28}
	\frac{\beta_{\rm{i}}}{1-\beta_{\rm{i}}} 
	\left( \frac{m_{{\rm Pl}}}{M} \right)^{1/2} \,.
\end{equation}
so that the constraint $\Omega_{\rm{pbh,0}} < 1$ leads to
\begin{equation}
\frac{\beta_{\rm{i}}}{1-\beta_{\rm{i}}}< 1.5 \times 10^{-29} \left( 
	\frac{M}{m_{{\rm Pl}}} \right)^{1/2} \,.
\end{equation} 
 
The calculation for the relic limit can be carried out identically
leading to
\begin{equation}
\frac{\beta_{\rm{i}}}{1-\beta_{\rm{i}}}< 1.5 \times 10^{-29} \left( 
	\frac{M}{m_{{\rm Pl}}} \right)^{3/2} \,.
\end{equation}
with the extra factor of $(M/m_{{\rm Pl}})$ from Eq.~(\ref{rel}).

\begin{figure}[t]
\centering 
\leavevmode\epsfysize=6cm \epsfbox{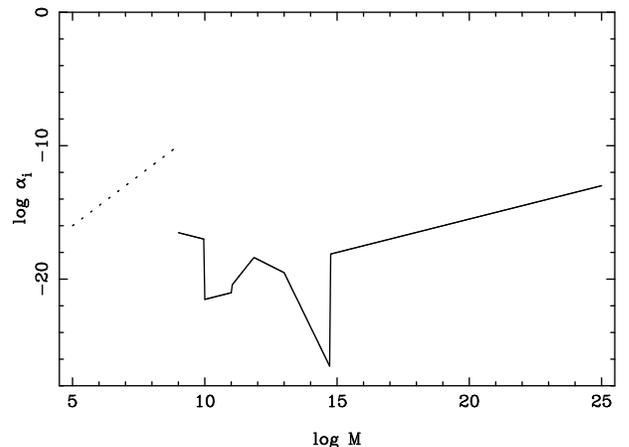}\\ 
\caption[fstand]{\label{fstand}  The tightest limits on the initial
mass fraction of PBHs, $\alpha_{\rm_{i}}$. The relic constraint is shown 
as a dotted line, emphasizing that it is not compulsory.}
\end{figure}

The various limits on the initial mass fraction of PBHs are displayed
in Table~\ref{initmass} and illustrated in Fig.~\ref{fstand}. 

\begin{figure}[t]
\centering 
\leavevmode\epsfysize=8cm \epsfbox{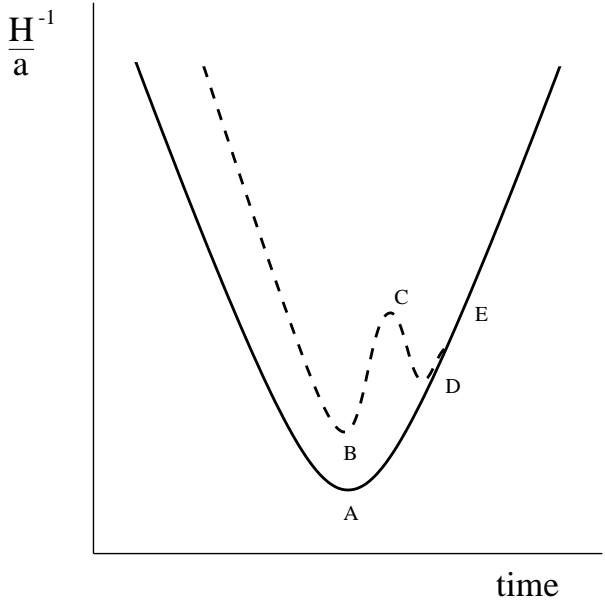}\\ 
\vspace*{0.1cm}
\caption[fig2]{\label{fig2} 
A schematic of the variation of the comoving Hubble radius
$(H^{-1}/a)$ with time for the standard evolution of the universe
(solid line) and with thermal inflation (dashed line). Points A and B
correspond to the end of the original period of inflation in the
standard evolution and with thermal inflation respectively. Thermal
inflation begins at point C and finishes at D, after which time the
comoving Hubble radii must coincide. Between D and E the scales which
are entering the Hubble radius are doing so for the second time so
that no PBHs are formed in this region. The values of the comoving
Hubble radius and the horizon mass at these points are displayed in
Table~\ref{comhub}. We denote the current comoving Hubble radius and
horizon mass as $(H^{-1}/a)_0$ and $M_{\rm{H0}}$ respectively.}
\end{figure}

\subsection{With a period of thermal inflation}    
                                                                               
We model the period of thermal inflation by assuming that at
$T= 10^{7}$~GeV the energy density in radiation splits into two
components, with one degree of freedom becoming the inflaton and the
remainder staying as radiation:
\begin{equation}
 \rho_{\rm{rad}} \rightarrow \tilde{\rho}_{\rm_{rad}} + \rho_{ \phi} \,,
\end{equation}
where $\rho_{\phi} = \pi^2 (10^7 \, {\rm GeV})^4/30$ is the false
 energy density of the flaton field which drives
thermal inflation, i.e.
\begin{equation}
\frac{\pi^2}{30} g_{\star} T^4 \rightarrow  
	\frac{\pi^2}{30} (g_{\star}-1) T^4 + \rho_{\phi} \,.
\end{equation}
Thermal inflation proper then commences once the radiation has
redshifted sufficiently, at 
\begin{equation}
T_{\rm{ti}} = \frac{10^{7} \rm {GeV}}{(g_{\star}-1)^{1/4}} \,,
\end{equation}
when $\rho_{\phi} > \tilde{\rho}_{\rm{rad}}$ and continues until the
flaton field rolls to its true vacuum state at $T = 10^3$~GeV. We then
assume that reheating is efficient so that the universe is reheated to
$T_{\rm{ti}}$ and the subsequent evolution has its standard form. 
Inefficient reheating, normally modelled as matter
domination, would make little difference. The duration of thermal
inflation is negligible in terms of the evaporation time-scale, so its
dominant effect on PBH evolution is to dilute their energy density
relative to that of radiation. The radiation energy density is the
same before and after thermal inflation (assuming efficient reheating)
whilst the energy density of PBHs is diluted by a factor $
\rho_{\rm{i}} / \rho_{\rm{f}} = (a_{\rm{f}}/a_{\rm{i}})^{3} \sim
(10^{4})^3$, where $a_{{\rm i}}$ and $a_{{\rm f}}$ are the scale
factors immediately before and after thermal inflation.
 
Fig.~\ref{fig2} illustrates the variation of the comoving Hubble radius 
both with and without thermal inflation,
with the Hubble radius and $M_{\rm{H}}$ at important points being
given in Table~\ref{comhub}.

Provided that the PBHs do not come to dominate before thermal
inflation, the limits on the initial black hole mass fraction from the
evaporation constraints are simply weakened by a factor of
$10^{12}$. The condition for the universe to be radiation dominated
before thermal inflation commences is
\begin{equation}
\left( \frac{\beta}{1-\beta} \right)_{\rm {T_{ti}}} = 
	\frac{\beta_{\rm{i}}}{1 -\beta_{\rm {i}}} \left( 
	\frac{T_{\rm {form}}}{T_{\rm{ti}}} \right) <1 \,.
\end{equation}
Using Eq.~(\ref{M}) this requires
\begin{equation}
\label{dom}
\frac{\beta_{\rm{i}}}{1 -\beta_{\rm{i}}} < 6 \times 10^{-12} \left(
	\frac{M}{m_{{\rm Pl}}} \right)^{1/2} \,.
\end{equation}

If the PBHs come to dominate at $T>10^{7}$~GeV 
then thermal inflation can only commence once the false energy density
of the flaton field dominates the energy density on PBHs, $
\rho_{\phi} > \rho_{\rm{pbh}} $. This delays the start of inflation,
so that a smaller number of $e$-foldings of inflation occur; however, the
dilution of the PBH energy density relative to that of radiation remains
the same. During the intermediate period $\rho_{\phi}$ remains constant
while $\rho_{\rm{pbh}} \propto a^{-3}$, so $\rho_{\rm{pbh}}$ is rapidly
reduced to below $\rho_{\phi}$ and the resulting constraints from 
Hawking radiation on the initial mass fraction are only slightly tighter
than if thermal inflation commences at $T_{{\rm ti}} = 10^{7}/3$~GeV.
However these limits are tighter than Eq.~(\ref{dom}), so that in fact 
PBHs with $M>10^{9}$~g cannot be produced with sufficient abundance that
they come to dominate the universe before thermal inflation.

\begin{table}[t]
\begin{center}
\begin{tabular}{|c|c|c|} 
Point & Comoving Hubble Radius & Horizon Mass \\ 
\hline 
A & $\left( \frac{H^{-1}}{a} \right)_0 \frac{T_{0}}{T_{{\rm RH}}}$ & 
$ M_{{\rm H,0}} \left( \frac{T_{0}}{T_{{\rm RH}}} \right)^2 $ \\ 
\hline 
B & $ 10^{4} \left( \frac{H^{-1}}{a} \right)_0 \frac{T_{0}}{T_{{\rm RH}}} 
$
& $ M_{{\rm H,0}} \left( \frac{T_{0}}{T_{{\rm RH}}} \right)^2 $ \\
\hline 
C & $\left( \frac{H^{-1}}{a} \right)_0 \frac{T_{0}}{10^{3}
\rm{GeV}}$ & $ M_{{\rm H,0}} \left( \frac{T_{0}}{10^{7} \rm{GeV}}
\right)^2 $ \\ 
\hline 
D & $\left( \frac{H^{-1}}{a} \right)_0
\frac{T_{0}}{10^{7} {\rm GeV}}$ & $ M_{{\rm H,0}} \left(
\frac{T_{0}}{10^{7} \rm{GeV}} \right)^2 $ \\ 
\hline 
E & $\left(
\frac{H^{-1}}{a} \right)_0 \frac{T_{0}}{10^{3} \rm{GeV}}$ & $ M_{{\rm
H,0}} \left( \frac{T_{0}}{10^{3} \rm{GeV}} \right)^2 $ 
\end{tabular}
\end{center}
\caption[comhub]{\label{comhub} Comoving Hubble Radii and horizon
masses at points on Fig.~\ref{fig2}.}
\end{table}

In the case of lighter PBHs, which are only constrained by the 
present-day relic density, the condition for radiation
domination before thermal inflation commences, Eq.~(\ref{dom}), is more 
constraining than the requirement that $\Omega_{\rm{rel,0}}<1$. These 
light PBHs can therefore come to dominate before thermal inflation
and delay its start, as discussed above, although the resulting
constraint from $\Omega_{\rm{rel,0}}<1$ is virtually the same as when
the universe is radiation dominated at the start of thermal inflation.
 
PBHs with mass $M<10^{18}$~g are formed before thermal inflation and
their energy densities will therefore be diluted by thermal inflation,
so that the gravitational constraint for $ 10^{15} \, {\rm g} < M <
10^{18} \, {\rm g}$ and the relic constraint will be weakened by a
factor $10^{12}$. This allows PBHs with $M<10^9$~g to be produced
with initial abundance $\beta_{{\rm i}}$ close to one, although not
arbitrarily so since sufficient thermal inflation to dilute the present
day relic density must occur. During inflation $M_{\rm{H}}$ remains 
constant
before increasing as $T^{-2}$ again after inflation; however, until the
temperature falls to $T=10^{3}$~GeV once more the scales that are
entering the Hubble radius will be doing so for the second time having
first entered at $T> T_{{\rm ti}}$ (and possibly forming black holes)
before being inflated away again. There are therefore no new density
perturbations present to collapse into PBHs, so whilst $T$ falls from
$10^{7}$~GeV to $10^{3}$~GeV after thermal inflation no PBHs form leading 
to a `missing' mass
range $10^{18} \, {\rm g} < M <10^{26} \, {\rm g}$. The gravitational
constraints on PBHs with mass $M>10^{26}$~g, which form after thermal
inflation, are unchanged.  

\begin{figure}[t]
\centering 
\leavevmode\epsfysize=6cm \epsfbox{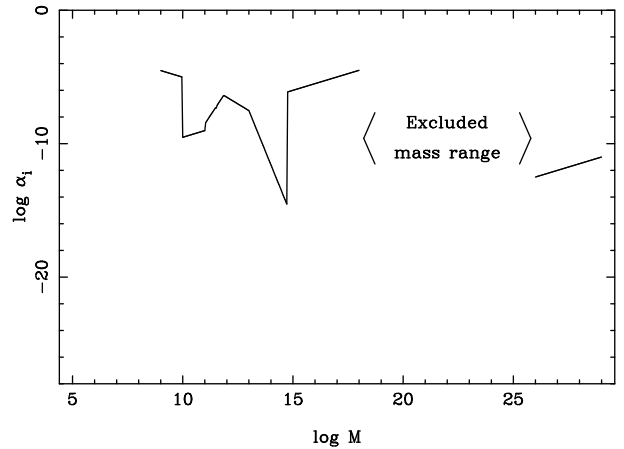}\\ 
\caption[ftherm]{\label{ftherm} The tightest limits on the initial mass
fraction of PBHs $\alpha_{\rm{i}}$ if thermal inflation occurs, on the 
same 
vertical scale as Fig.~\ref{fstand}. The
gap $10^{18} \, {\rm g} <M<10^{26} \, {\rm g}$ is the excluded mass
range. For $M<10^{9} \rm{g}$ a large initial mass fraction $\sim 1$
of PBHs is allowed.}
\end{figure}

The various limits on the initial mass fraction of PBHs are displayed
in Table~\ref{ttherm} and illustrated in Fig.~\ref{ftherm}. 

\begin{table}[t]
\begin{center}
\begin{tabular}{|c|c|} 
Constraint & Range \\ \hline $ \alpha_{\rm{i}}< 3\times10^{-4} \,
(10^{9} \, {\rm g}/M)$ & $ 10^{9} \, {\rm g}< M < 10^{13} \, {\rm g}$ \\ 
\hline $ \alpha_{\rm{i}}
<3\times 10^{-15}$ & $2\times10^{14} \, {\rm g}<M<5\times10^{14}
\, {\rm g}$ \\ \hline $
\alpha_{\rm{i}}<3\times 10^{-5} \,
(10^{9} \, {\rm g}/M)^{1/2} $ & $10^{9} \, {\rm g}
< M < 10^{11} \, {\rm g}$ \\ 
\hline $\alpha_{\rm{i}} <3\times10^{-10} \,
(M/10^{10} \, {\rm g})^{1/2} $ & $10^{10} \, {\rm g}< M < 10^{11} \, 
{\rm g}$ \\ \hline $
\alpha_{\rm{i}} <3\times 10^{-9} \,
(M/10^{9} \, {\rm g})^{5/2} $ & $10^{11} \, {\rm g}< M <10^{13} \, {\rm 
g}$ \\ \hline $
\alpha_{\rm{i}} < 1 \times 10^{11} \,
(M/10^{15} \, {\rm g})^{3/2} $ & $M<10^{15} \, {\rm g}$ \\ \hline $
\alpha_{\rm{i}} < 1\times10^{-7} \,
(M/10^{15} \, {\rm g})^{1/2} $ & $10^{15} \, {\rm g} < M <10^{18} \, 
{\rm g}$ \\ \hline $
\alpha_{\rm{i}} < 1\times10^{-19} \,
(M/10^{15} \, {\rm g})^{1/2} $ & $M>10^{26} \, {\rm g}$ \\ 
\end{tabular}
\end{center}
\caption[limin]{\label{ttherm} Limits on initial mass fraction of PBHs if 
thermal inflation occurs}
\end{table}

\section{Limits on the spectral index}

\label{s:index}

Using Eq.~(\ref{sigma}), the limits on the initial mass fraction can
be used to constrain the spectral index of the density perturbations
$n$. The four-year fitting function to the COBE data, assuming negligible
contribution from gravitational waves, gives the normalization at the
present Hubble scale ($ k = a_{\rm{0}} H_{\rm{0}}$) \cite{BLW:COBEn}
\begin{equation}
\delta_{\rm{H}}(\rm{n}) = 1.91\times10^{-5} \exp{\left[
	1.01(1-\rm{n})\right]} \,.
\end{equation}

When inserted in Eq.~(\ref{sigmam}) and numerically integrated, this
allows us to normalize $\sigma(M)$ for a chosen fixed $M$. We take the
COBE data to correspond to a scale equal to the present Hubble radius
$R_{0} = 3000 h^{-1}$, so that $M_{0}= 1 \times 10^{56}$~g is the
present horizon mass which gives a normalization $\sigma(10^{56}
\rm{g}) = 9.5\times 10^{-5}$. To the level of accuracy at which we are
working $\sigma(10^{56}\rm{g})$ varies only slowly with $n$.
Putting this in Eq.~(\ref{sigma}) gives
\begin{equation}
\sigma_{{\rm hor}}(M) = 9.5 \times 10^{-5} \, \left( 
	\frac{M}{10^{56} \, {\rm g}}\right)^{(1-n)/4} \,,
\end{equation}
Here we ignore the change in slope to $(1-n)/6$ which occurs at matter 
domination (corresponding to a mass $M_0 \, t_{{\rm eq}}/t_0 \simeq 
10^{49}$~g), which only changes the constraint on $n$ by about 0.01.

Taking logarithms in Eq.~(\ref{beta}) gives
\begin{equation}
\sigma(M) = 0.15 \left( \log_{10}\sigma(M) - \log_{10} {\beta_{i}} 
	\right)^{-1/2} \,.
\end{equation}
Since $\beta_{\rm_{i}}/( 1 -\beta_{\rm{i}}) \ll 1$, we can take
$\beta_{\rm_{i}} \approx \beta_{\rm {i}}/ ( 1 - \beta_{\rm{i}})$ and
find the upper bounds on $n$ corresponding to each of the limits found
in the previous section.

\subsection{Standard evolution of the Universe}

\begin{figure}
\centering 
\leavevmode\epsfysize=6cm \epsfbox{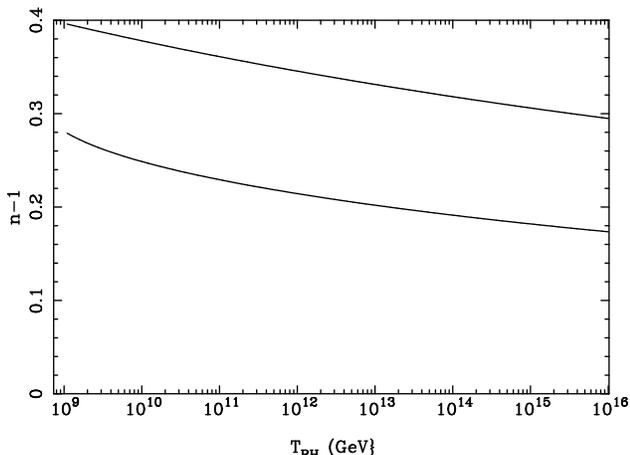}\\ 
\caption[fig4]{\label{fig4} The variation of the limits on $n$ with 
reheating temperature from the relic constraint (lower line) and from
$\delta(M_{\rm{min}})<1$.}
\end{figure}

The appropriate constraint is normally determined by the lightest PBHs 
that 
can form, given by Eq.~(\ref{mhor}). The tightest limit is $n < 1.22$ from 
the deuterium destruction
constraint evaluated at $M \sim10^{11}$~g, although all the
constraints due to the evaporation of the PBHs require $n<1.24$. The
tightest constraint from the limit on the present density of PBHs is
$n<1.31$ at $M\sim 10^{15}$~g, weakening with increasing $M$. The
relic constraint may place a stronger constraint on $n$, depending on
the minimum mass of PBHs produced, which is determined by the
reheating temperature $T_{\rm{RH}}$ after the initial period of
conventional inflation
\begin{equation}
M_{\rm{min}} = M_{0} \left( \frac{T_{0}}{T_{\rm{RH}}} \right)^2 \,.
\end{equation}
If $T_{\rm{RH}}<10^{9}$~GeV, there is not sufficient time for
$\rho_{\rm{PBH}}$ (and after evaporation $\rho_{\rm{rel}}$) to
increase by enough relative to $\rho_{\rm{rad}}$ for the relics to
dominate today. For comparison, in Fig.~\ref{fig4} we plot the
variation of the limit on $n$ with reheating temperature and the limit
from the simple requirement that $\delta(M_{\rm{min}}) <1$, where
$M_{\rm{min}} = (T_{\rm{Pl}}/T_{\rm{RH}})^2 m_{\rm_{Pl}}$ is the mass
of the lightest PBHs formed immediately after the first period of
inflation.

In summary, the tightest Hawking radiation constraint on $n$ we obtain
is $n<1.22$, although if the reheating temperature is sufficiently
high the existence of relics may lead to a tighter limit. In
Ref.~\cite{CGL:PBH} a much weaker limit of $n<1.48$ from Hawking radiation
constraints was found, which becomes tighter if relics are formed
($n<1.4$ if $T_{{\rm RH}} = 10^{16}$~GeV). Our stronger constraint arises 
from our correcting of two errors in their paper, which both go the same 
way. The first is that we use the correct scaling law $\sigma(M) \propto
M^{(1-n)/4}$ for the variance at the horizon scale during 
radiation-domination, as mentioned earlier. The second is that our 
normalization to COBE is much higher. They omitted a numerical prefactor 
$\sqrt{512\pi/75}$ in the power spectrum expression, and also assumed that 
the normalization of $\delta_{{\rm H}}(k)$ (which from the four-year 
COBE data is $2 \times 10^{-5}$) and $\sigma(M)$ were 
interchangeable. In combination this raises the COBE normalization by a 
factor of over twenty. A much smaller additional correction is that the 
normalization from the COBE four-year data~\cite{G:Cnew} is higher than  
that from the first year's data~\cite{SB:Cold}.  

\subsection{With a period of thermal inflation}

During thermal inflation the comoving Hubble radius $(H^{-1}/a)$
varies as $T$ so Eq.~(\ref{hub}), for when a given comoving scale $k$
crosses the Hubble radius, becomes
\begin{equation}
k^{-1} = \left( \frac{H^{-1}}{a} \right) = \left( \frac{H^{-1}}{a}
	\right)_{{\rm eq}} \left( \frac{T_{{\rm eq}}}{T} \right) \left(
	\frac{T_{\rm{ti}}}{10^{3} \rm{GeV}} \right) \,,
\end{equation}
whilst the relation between horizon mass and temperature remains
unchanged, since during inflation the horizon mass is constant so that
for PBHs formed in between the two periods of inflation a given mass
PBH will correspond to a larger scale than in the standard
scenario. This leads to the modification of Eq.~(\ref{sigma}):
\begin{equation}
\sigma_{{\rm hor}}(M) = \sigma_{{\rm hor}}(M_{{\rm eq}}) \left[
	\frac{M}{M_{{\rm eq}}} \left( \frac{10^3 \, \rm{GeV}}{T_{\rm{ti}}}
	\right)^{2} \right]^{(1-n)/4} \,.
\end{equation}

In this case the tightest constraint is $n<1.29$ from the deuterium
destruction constraint evaluated at $M \sim 10^{10}$~g, with all the
constraints due to evaporation requiring $n<1.34$. In this case the
constraints from gravitation and relics are weaker. The limit from the
present-day density leads to $n<1.35$ for $M\sim 10^{15}$~g PBHs,
which are diluted by the thermal inflation, and $n<1.49$ for
$M\sim10^{26}$~g PBHs, which are the lightest formed after thermal
inflation.  For $T_{\rm{RH}}<10^{14}$~GeV the relics do not constrain
$n$, since even if $\beta_{\rm{i}}$ is close to one the relics will be
diluted away. However, for $T_{\rm{RH}} = 10^{16}$~GeV we find
$n<1.3$ is required although this limit rapidly weakens as the reheat
temperature falls towards $10^{14}$~GeV.

\subsection{The effect of non-gaussianity}

\label{nongauss}

We now return to the issue of the gaussianity assumption used to obtain 
Eq.~(\ref{gauss}). Bullock and Primack \cite{BP:gauss} have stressed that 
the normal justification of gaussianity relies on the perturbations being 
very small, something which can no longer be justified when considering 
PBH 
formation. Unfortunately, the non-gaussian correction is strongly model 
dependent, and in detail must be examined case-by-case. They 
numerically study three different `toy' models, in one case finding 
negligible non-gaussianity but in the other two finding a very significant 
suppression in the number of large perturbations, which are of course 
exactly those utilized for PBH formation.

Expressed in terms of the probability of high density perturbations, the 
suppression can be very dramatic; in one of their toy models the 6-sigma 
perturbations are suppressed by a factor of $10^{150}$! However, despite 
that the effect on the constraint on $n$ is not large, because that asks a 
rather different question, namely how much larger does the variance 
$\sigma(M)$ have to be so that the non-gaussian perturbations reproduce 
the number density of gaussian ones? In their most extreme example, the 
answer is about three times; the perturbations with the appropriate number 
density correspond to about 3-sigma perturbations in the non-gaussian case 
rather than the 9-sigma or so perturbations of the gaussian case 
\cite{BP:gauss}.

The conclusion then is that non-gaussian effects are model-dependent, and 
in the worst tested case weaken the constraints on $\sigma_{{\rm hor}}(M)$ 
by about a factor three. Non-gaussianity clearly cannot do much more than 
this, as the low required number density keeps us to the tail regardless 
of the amount of non-gaussianity. {}From Eq.~(\ref{sigma}), using the COBE 
normalization to keep $\sigma_{{\rm hor}}(M_0)$ fixed, this weakens the 
constraint on $n$, in the worst case of non-gaussianity, by about 0.05. 

\section{Conclusions}

The interpretation of constraints on primordial black holes depends
sensitively on the whole history of the universe from their time of
formation. We re-examined the constraints assuming the standard 
radiation-dominated cosmology, and by correcting two errors in Carr et 
al.~\cite{CGL:PBH} found a significantly tighter constraint on the 
spectral index than they did, $n \lesssim 1.25$. This is in fact presently 
the tightest constraint on the spectral index, being somewhat stronger 
than large-scale structure constraints in the most general cosmologies, 
and much stronger than the constraint from distortions to the microwave 
background spectrum \cite{HSS}. On the other hand, its application 
requires a belief that the spectral index remains constant over a much 
wider range of scales than the others, which is certainly possible but not 
mandatory. In general circumstances one must impose the general 
constraints on the formation rate we derived in Section~\ref{s:massfrac}.

We have analyzed the changes to the standard scenario
brought about if a period of thermal inflation takes place in the
early universe after the black holes form. Thermal inflation leads to
a significant weakening on the constraint on the density perturbation
spectrum. In Section~\ref{s:massfrac} we recomputed the constraints on the 
initial mass fraction of black holes; especially at low masses the 
constraints become very weak indeed. In Section~\ref{s:index}, we assumed 
a power-law spectrum of perturbations and constrained the spectral index 
$n$; we found the constraint weakened (relative to the standard cosmology) 
to $n \lesssim 1.3$. A novel
additional feature is that thermal inflation predicts a missing mass
range for black holes, extending up from $10^{18}$~g to $10^{26}$
g. It will be hard to probe this range as black holes of these
masses have negligible evaporation. However, if for some reason
thermal inflation can start at a higher energy than currently supposed,
say $10^{10}$~GeV, then the missing mass range could extend down into
the evaporating regime.

It seems quite likely that density perturbations large enough to form PBHs 
will exhibit a significantly non-gaussian probability distribution, as 
emphasized by Bullock and Primack \cite{BP:gauss}. However, we have shown 
that this does not much alter the constraints; in the worst case 
it corresponds to a weakening 
of about 0.05 (regardless of whether or not thermal inflation 
occurred) which is comparable to the change due to thermal inflation. 
But typically the correction is expected to be quite a bit smaller 
than this \cite{BP:gauss}, and hence smaller than the uncertainty 
in the cosmological model at early times.

\section*{Acknowledgments}

A.M.G.~was supported by PPARC and A.R.L.~by the Royal
Society. A.R.L.~thanks John Webb and UNSW for hospitality while some of 
this work was carried out. We thank Bernard Carr and Jim Lidsey for their 
comments, and Pedro Viana for useful discussions and
providing the code for evaluation of $\sigma(M)$. We acknowledge use
of the Starlink computer system at the University of Sussex.

\end{document}